\documentstyle[prl,aps,multicol,epsfig]{revtex}
\begin{document}
\draft
\title{Coulomb Drag for Strongly Localized Electrons: Pumping
Mechanism}
\author{M.E.\ Raikh$^{1}$ and Felix von Oppen$^{2}$}
\address{
$^1$ Department of Physics, University of Utah, Salt Lake City,
Utah 84112, U.S.A.\\
$^2$ Institut f\"ur Theoretische Physik,
Freie Universit\"at Berlin, Arnimallee 14,
14195 Berlin, Germany
}
\date{\today}
\maketitle
\tighten
\begin{abstract}
  The mutual influence of two layers with strongly localized electrons
  is exercised though the random Coulomb shifts of site energies in
  one layer caused by electron hops in the other layer. We trace how
  these shifts give rise to a voltage drop in the passive layer, when
  a current is passed through the active layer. We find that the
  microscopic origin of drag lies in the {\em time correlations} of
  the occupation numbers of the sites involved in a hop. These
  correlations are neglected within the conventional Miller-Abrahams
  scheme for calculating the hopping resistance.
\end{abstract}

\begin{multicols}{2}

Since the subject was first introduced in pioneering papers
\cite{pogrebinskii77,price83}, Coulomb drag between two parallel
electronic layers has commanded a lot of attention.  Stimulated by the
early experiments \cite{gramila91,sivan92}, the development of the
theory proceeded in two directions:

\noindent({\em i}) progress on the formalism for calculating 
drag between conventional two-dimensional electron gases
\cite{tso92,jauho93,zheng93,flensberg94,kamenev95,flensberg95,flensberg'95}.

\noindent({\em ii}) accomodation of various
realizations of interacting layers.  These include electron-hole
layers \cite{tso93,swierkowski95,vignale96}, layers in the
superconducting state \cite{kamenev95}, electronic layers with
tunneling links \cite{oreg98}, strongly disordered layers at
the onset of Anderson localization \cite{shimshoni97}, diffusive
layers with correlated disorder \cite{gornyi99}, and double-layer
systems in a perpendicular magnetic field.  In the latter case, the
picture of drag depends on the magnetic-field regime, namely
classically strong fields \cite{khaetskii98}, quantizing fields 
\cite{bonsager96,vonOppen01}, the
vicinity of the integer quantum Hall transition \cite{shimshoni94},
and the fractional quantum Hall regime
\cite{ussishkin97,sakhi97,kim99,narozhny01}.

For all these interacting two-dimensional systems, the theories of
Coulomb drag shared the common scenario of quasiparticles (electrons,
holes, or composite fermions
\cite{ussishkin97,sakhi97,kim99,narozhny01}) in two contacting layers
scattering off one another in the course of ballistic motion,
diffusion \cite{zheng93} or anomalous diffusion \cite{shimshoni94}.
This scattering results in a non-zero average momentum transfer
between the active (current-carrying) and passive (open-circuit)
layers. A voltage drop is then induced across the passive layer to
ensure the absense of a net momentum.

Consider now the deeply insulating regime, where the localization
radius of electronic states is smaller than the inter-electronic
distance. Obviously, the momentum is not a good quantum number in
this case, so that the conventional scenario of Coulomb drag does {\it
  not} apply. In addition, the picture of long-range time fluctuations
of the electron density, and thus the language of {\em spatially
  averaged} response functions, is inadequate in the strongly
localized regime. This is because the electron motion is due to
hopping, which is characterized by an exponentially wide spread in the
{\em local} transition frequencies.

Instead of momentum exchange due to collisions, the coupling between
the contacting strongly localized systems is based on 
random shifts of energy levels in the passive layer caused by electron
hops in the active layer, and vice versa. In this situation, it is not
immediately obvious how these shifts ``communicate'' the overall
direction of current from the active to the passive layer.

\begin{figure}
\begin{center}
\epsfig{figure=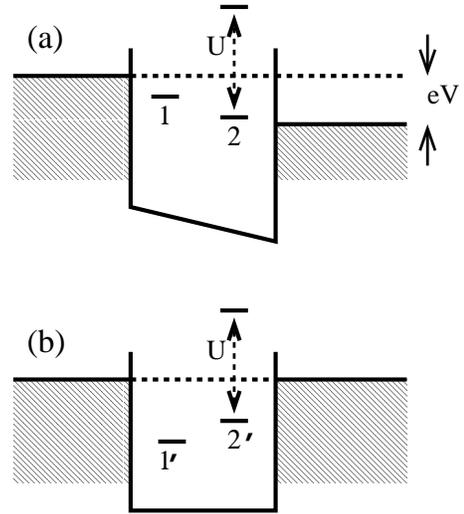,width=6cm}
\end{center}
\caption{Schematic drawing of (a) active layer with applied voltage $V$
  and two localized states $1$ and $2$.  $U$ indicates the Coulomb
  energy by which level $1$ ($2$) is elevated if level $2$ ($1$) is
  occupied. (b) The passive layer also involves two localized states
  $1'$ and $2'$ , but has no voltage applied.}
\end{figure}

This question is addressed in the present paper. We find that Coulomb
drag in the strongly localized regime is due to the {\em correlated
  character} of hopping transport. To clarify this point, consider two
neighboring sites, $1$ and $2$, which belong to the current-carrying
cluster \cite{book}. At nonzero temperature the time evolution,
$n_1(t)$ and $n_2(t)$, of the occupation numbers of sites has the form
of telegraph noise.  In equilibrium, the average
$K_{12}(\tau)=\left\langle n_1(t)n_2(t+\tau)\right\rangle $ is an {\em
  even} function of $\tau$. Suppose now that current flows in the
direction $1\rightarrow 2$. Then we have $K_{12}(\tau)>K_{12}(-\tau)$
for $\tau > 0$. This reflects the fact that, as the hops within a pair
of sites occur preferentially from $1$ to $2$, the occupation numbers
change {\em in a certain sequence}. This asymmetry is the analog of
the current-induced asymmetry between wave vectors ${\bbox q}$ and
$-{\bbox q}$ of the thermal density fluctuations in the metallic
regime, and thus, it is responsible for Coulomb drag between layers
with strongly localized electrons.  

Note in passing, that conventional theories of hopping transport
neglect the asymmetry in $K_{12}(\tau)$.  In both the non-interacting
\cite{book} and the interacting \cite{review} cases, the resistance of
an elementary hop is computed under the assumption of uncorrelated
occupation numbers, {\em i.e.}  $\left\langle
  n_1(t)n_2(t^{\prime})\right\rangle =\left\langle n_1 \right\rangle
\left\langle n_2 \right\rangle $.

\begin{figure}
\centerline{
\epsfxsize=6cm
\epsfbox{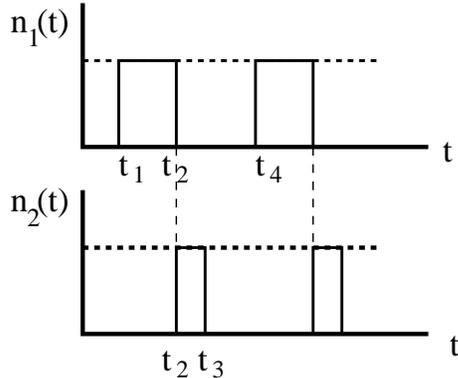}
}
\caption{Schematic time evolution of occupations $n_1(t)$ and $n_2(t)$
of
levels $1$ and $2$, respectively, in the active layer under conditions
of a finite current. }
\end{figure}

To show how the preferential sequence in the change of the occupation
numbers transforms into drag we consider the simplest possible model
as illustrated in Fig.\ 1. Within this model, the active and the
passive layer are each represented by a pair of sites ($1$ and $2$)
coupled to metallic contacts ($l$ and $r$).  Both ``active'' and
``passive'' pairs of sites can be either empty or singly occupied. The
corresponding conditions are $\varepsilon_{1,2} + U > E_F+ eV$ for the
active layer, where $U$ is the Coulomb interaction between two
electrons on sites $1$ and $2$ and $V$ the applied voltage, and
$\varepsilon_{1^{\prime},2^{\prime}} + U > E_F$ for the passive layer.
Due to this condition an electron occupying, say, site $2$ elevates
the energy level of site $1$ above the Fermi level, thus forbidding
the tunneling process $l\rightarrow 1$ (see Fig.\ 1). We will treat
this model first in the strongly-nonlinear regime, in which the 
temperature $T$ is much lower than $V$. Subsequently, we will consider 
the Ohmic regime, where $V\ll T$. 

In the strongly-nonlinear regime, we can neglect all activation
processes and thus, only a {\em single} sequence of hops is possible
in the active layer. Within this sequence, the occupation numbers
$n_1$ and $n_2$ of sites $1$ and $2$ undergo the transformations
$(1,0)\rightarrow (0,1)\rightarrow (0,0)\rightarrow (1,0)$.  With each
repeated cycle of this sequence, an electron is transferred from the
left to the right contact.  The time evolution of $n_1$ and $n_2$
during one cycle is illustrated in Fig.\ 2. The strong asymmetry in
$K_{12}(\tau)$ is evident. The average current through the active
layer is equal to \cite{raikh92,kinkhabwala00} $\left\langle I_a\right\rangle
=e/(\tau_1+\tau_2+\tau_3)$, where $\tau_1$, $\tau_2$ and $\tau_3$ are
the average waiting times for the transitions $l\rightarrow 1$,
$1\rightarrow 2$ and $2\rightarrow r$, respectively. In the strongly
nonlinear regime, the current $\left\langle I_a\right\rangle $ is
independent of the voltage drop, $V$.

\begin{figure}
\centerline{
\epsfxsize=6cm
\epsfbox{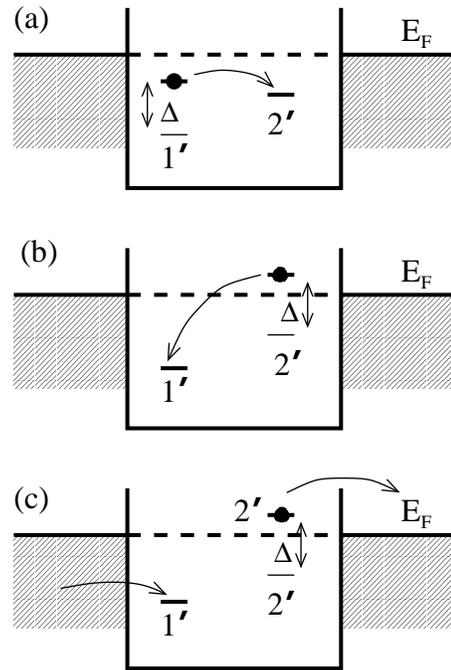}
}
\caption{Variants of evolution of the level occupations in the passive
layer during a single cycle in the active layer.}
\end{figure}

Consider now the response of the passive layer to a single cycle of
$\left(n_1(t),n_2(t)\right)$.  In the configuration $n_1=1$, $n_2=0$,
the occupied site $1$ elevates the level $\varepsilon_{1^{\prime}}$ by
some $W_1$, and the level $\varepsilon_{2^{\prime}}$ by some
$W_2=W_1-\Delta <W_1$. Conversely, when $n_1=0$, $n_2=1$, the level
$\varepsilon_{1^{\prime}}$ is elevated by $W_2$, while the level
$\varepsilon_{2^{\prime}}$ is elevated by $W_1$.  (We define the level
positions $\varepsilon_{1^{\prime}}$, $\varepsilon_{2^{\prime}}$ in
the passive layer with respect to an ``empty'' active layer,
$n_1=n_2=0$.) As we will see, the pair $1^{\prime}$, $2^{\prime}$ is
most ``sensitive'' when the conditions
\begin{eqnarray}
\label{first}
E_F >&& \varepsilon_{1^{\prime}}+W_1 > \varepsilon_{2^{\prime}} +W_2  \\
\label{second}
&&\varepsilon_{2^{\prime}}+W_1 > E_F
\end{eqnarray}
are met.  Indeed, by virtue of Eq.\ (\ref{first}), the passive layer
can only make the transition $1^{\prime}\rightarrow 2^{\prime}$ during
the interval $t_1<t<t_2$ when site $1$ in the active layer is
occupied. This is illustrated in Fig.\ 3a.  If the hop
$1^{\prime}\rightarrow 2^{\prime}$ does take place, then two
transitions within the passive layer become possible during the next
time interval $t_2<t<t_3$:

\noindent{\em 1})\ the electron on
site $2^{\prime}$ goes back to $1^{\prime}$, cf.\ Fig.\ 3b.
This transition is energetically favorable due to
Eq.\ (\ref{second}).

\noindent{\em 2})\ the electron tunnels from $2^{\prime}$
into the right contact, $r^{\prime}$, emptying the passive
layer.  Since $\varepsilon_{1^{\prime}}+ W_2 <E_F$, this opens
the possibility for the hop $l^{\prime}\rightarrow 1'$ (see
Fig.\ 3c).

In the first case, there is no transfer of charge from the left to the
right contact in the passive layer during the cycle in the active
layer. In the second case, however, provided that both transitions
$2^{\prime}\rightarrow r^{\prime}$ and $l^{\prime}\rightarrow
1^{\prime}$ took place during the interval $t_2<t<t_3$, the cycle in
the active layer results in the transfer of an electron from
$l^{\prime}$ to $r^{\prime}$. This transfer is nothing but the drag
current.

The way the drag current is induced here closely resembles the
operation of a classical {\em electron pump}
\cite{kouwenhoven91,pothier91,pothier92} where two phase-shifted rf
signals are applied either to the barriers \cite{kouwenhoven91} or to
the gate electrodes \cite{pothier91,pothier92} of a single or multiple
quantum dot structure. In contrast to adiabatic quantum pumping
\cite{altshuler99,switkes99}, classical pumping is due to the Coulomb
blockade, which forces an electron from one contact to enter the dot,
and then to leave the dot into the other contact during each rf cycle.
In the mechanism of hopping drag considered above, the pulses $n_1(t)$
and $n_2(t)$ play the role of the rf signals, their mutual phase shift
being governed by the direction of the current in the active layer.

We now turn to the calculation of the average drag current,
$\left\langle I_d\right\rangle$. Denote with ${\cal T}_1,{\cal T}_2$, 
and ${\cal T}_3$ the
waiting times corresponding to the transitions $l^{\prime}\rightarrow
1^{\prime}$, $1^{\prime} \rightarrow 2^{\prime}$, and
$2^{\prime}\rightarrow r^{\prime}$ in the passive layer.  The simple
system considered here mimics drag between two-dimensional hopping
layers when the transitions $1\rightarrow 2$ and
$1^{\prime}\rightarrow 2^{\prime}$ constitute ``bottlenecks'' for the
transport in active and passive layer, respectively, {\em i.e.}
$\tau_2$ and ${\cal T}_2$ are the ``long'' waiting times.  In this limit we
have $\left\langle I_a\right\rangle \approx e/\tau_2$. The calculation
of $\left\langle I_d\right\rangle$ is greatly simplified if the
``short'' waiting times are related as follows ${\cal T}_1 
\ll {\cal T}_3 \ll \tau_3$.  
The latter conditions ensure that the passive layer returns
to the ground state after each cycle in the active layer.  Indeed, if
the transition $1^{\prime}\rightarrow 2^{\prime}$ took place during
the time interval $t_1 < t < t_2$ (the corresponding probability is
equal to
$p_{1^{\prime}2^{\prime}}=1-\exp\left[-(t_2-t_1)/{\cal T}_2\right]$, 
then
the probability that, during the subsequent interval $t_2 < t < t_3$,
{\em both} transitions $2^{\prime}\rightarrow r^{\prime}$ and
$l^{\prime}\rightarrow 1^{\prime}$ take place is close to $1$.  This
is because the characteristic duration of this interval $(t_3-t_2)\sim
\tau_3$ is much longer than ${\cal T}_3$ and ${\cal T}_1$.  
In principle, the
transition $2^{\prime}\rightarrow r^{\prime}$ opens the possibility
for two follow-up hops, namely, $l^{\prime}\rightarrow 1^{\prime}$ and
$r^{\prime}\rightarrow 2^{\prime}$.  However, since 
${\cal T}_1 \ll {\cal T}_3$,
tunneling of an electron from the left contact onto site $1^{\prime}$,
thereby returning the passive layer into its ground state, occurs {\em
  before} the back hop $r^{\prime}\rightarrow 2^{\prime}$.  Once the
transition $l^{\prime}\rightarrow 1^{\prime}$ occurred, the passive
layer remains in the ground state until the end of the cycle in the
active layer $t=t_4$ (see Fig. 2). Thus, calculating the drag current
reduces to averaging $p_{1^{\prime}2^{\prime}}$ which yields
\begin{equation}
\label{id}
\left\langle I_d\right\rangle = \frac{\tau_2}{\tau_2 + {\cal T}_2}
\left\langle I_a\right\rangle = \frac{e}{\tau_2 + {\cal T}_2}.
\end{equation}
Remarkably, the drag current is {\em not} exponentially small compared
to the current in the active layer. Instead, both currents are of
{\em comparable magnitude}. 
This result is clearly very different from naive
predictions based on spatially-averaged response functions such as a
widely used Fermi-golden-rule-type expression for the drag
resisitivity \cite{zheng93}.

In principle, hops in the 
passive layer might, in turn, affect transport in the active layer.
In particular, occupation of site $2^{\prime}$ shifts level $2$
upward. As a result, the hop $2 \rightarrow 1$, in the direction
opposite to the net current, might become energetically favorable.
The condition that such ``feedback'' does not occur is $\varepsilon_1
> \varepsilon_2 + \Delta$.  As a concluding remark on our simplified
model, we note that for a given polarity of voltage across the active
layer (see Fig.\ 1) the activationless drag exists only if
$\varepsilon_1^{\prime}<\varepsilon_2^{\prime}$.

We now turn to the linear regime $V\rightarrow 0$ where transport in
the active layer becomes more complicated in two respects. Firstly,
charge-transfer processes become possible in {\it both} directions,
$l\rightarrow r$ and $r \rightarrow l$, with only a small difference
in their frequencies due to the applied voltage. Secondly, the
dynamics involves ``round-trip'' processes such as $l\rightarrow
1\rightarrow l$ or $l\rightarrow 1\rightarrow 2 \rightarrow 1
\rightarrow l$ that do not result in charge transfer between the
contacts.

It is important to note that the traditional description of hopping
transport \cite{book}, based on the Miller-Abrahams network, does not
capture the realistic occupation dynamics of the sites $1$ and $2$.
Indeed, in this description, a pair of sites is considered as an {\em
isolated system} so that the characteristic times for the changes in
the occupations $n_1$ and $n_2$ are long ($\sim \tau_2$ in our
notations). This is indeed the case for the non-ohmic regime
considered above where an electron {\em has} to spend a time $\sim
\tau_2$ at site $1$ before the transition $1 \rightarrow 2$ takes
place, since it cannot return to the contact $l$ without
activation. In the ohmic regime, however, while the waiting time,
$\tau_2$, is long, the occupation of sites $1$ and $2$, constituting a
``bottleneck'', changes many times between successive hops $1
\rightarrow 2$ (or $2\rightarrow 1$). In other words, a typical
charge-transfer process is preceded by many round-trip processes.

As in the non-ohmic regime, we restrict our considerations to a
particular domain of parameters for which the analysis of drag is
greatly simplified. We assume that (1) the energies of both sites in
the active layer are above the Fermi level
($\varepsilon_{1,2}>E_F$). Thus their average occupations are small,
$\left\langle n_1(t)\right\rangle, \left\langle n_2(t)\right\rangle
\ll 1$, i.e.  the states $1$ and $2$ are empty for long periods before
an electron enters for a time of the order of $\tau_1$ or $\tau_2$.
In addition, we assume for the passive layer that (2) the sites
$1^{\prime}$ and $2^{\prime}$ are symmetric in space, so that 
${\cal T}_1 \approx {\cal T}_3$, 
and close in energy (within temperature).  As a result,
unlike the situation in Fig.\ 3a, the site $1^{\prime}$ is also
elevated {\em above} the Fermi level, when the site $1$ in the active
layer is occupied. Finally, we demand that (3) the ``long'' 
time ${\cal T}_2$
in the passive layer is {\em shorter} than the typical time interval
during which the active layer is empty, but longer than the time when
one of sites $1$ or $2$ are occupied. The latter assumption implies
that an electron in the passive layer hops many times from
$1^{\prime}$ to $2^{\prime}$ and back between successive
round-trip processes in the active layer. We also demand 
${\cal T}_1,{\cal T}_2 \ll \tau_1,\tau_2$.

Under these conditions, we find a remarkable result, namely 
that the drag
current is simply half the current in the active layer. To understand
how this comes about, consider a single charge-transfer process in the
active layer, $l\rightarrow 1\rightarrow 2\rightarrow r$. Imagine
first that the passive layer is in the state $n_{1'}=0$, $n_{2'}=1$
when the transition $l\rightarrow 1$ takes place. In this case,
nothing happens in the passive layer until the transition
$1\rightarrow 2$ takes place, after which the passive layer will
undergo the transition $2'\rightarrow r'$ in essentially all cases.
This will be immediately followed by $l'\rightarrow 1'$. With the
transition $2\rightarrow r$ the system effectively returns to its
initial state and one electron has been transferred in {\it both}
layers. Since the passive layer has a 50\% chance of being in the
state $n_{1'}=0$, $n_{2'}=1$ at the outset, this implies
\begin{equation}
\label{result}
  \langle I_d\rangle = \frac{1}{2} \langle I_a\rangle.
\end{equation}
Eq. (\ref{result}) is based on the fact that there is {\em no}
comparable drag current when the passive layer initially has an
electron in $1'$. In this case, $l\rightarrow 1$ causes the
passive-layer transitions $1'\rightarrow l'$ and then $r'\rightarrow 2'$, 
which are reversed after $1\rightarrow 2$ occurs in the
active layer.
Using similar reasoning, it can be verified
that round-trip processes in the active layer do not lead to a
net drag current.

An actual sample in the deeply localized regime will consist of a
network of transresistors of the type described above.  Hence, we
should compute the transresistance of this network. This is by no
means a trivial task, even if the transresistances between elements of
the conducting networks in active and passive layer are known.  This
can be illustrated by considering the way in which transresistances
combine when connected in sequence or in parallel. These situations
are readily analyzed in terms of Kirchhoff's laws.  Two transresistors
in sequence have a transresistance equal to the sum of the individual
trans\-resistances,
\begin{equation}
  R_t^{(1+2)} = R_t^{(1)}+R_t^{(2)}
\label{series}
\end{equation}
similar to  ordinary resistors. Two transresistances
in parallel are less simple. Here, 
one can show that
\begin{equation}
   R_t^{(1\parallel2)} = {R_1R_1'R_t^{(2)} + R_2R_2'R_t^{(1)}\over
      (R_1+R_2)(R_1'+R_2')},
\label{parallel}
\end{equation}
where $R_i$ ($R'_i$) denote the resistances of the active (passive)
layer of the $i$-th transresistor. These expressions already allow one
to draw an important conclusion about the net transresistance of the
entire network. Even though the resistances $R_i$ and $R'_i$ are
exponentially large, adding transresistances in sequence or in
parallel does not lead to exponential changes in the transresistance.
In this sense, the drag current is of comparable magnitude as the
current in the active layer, even for the entire network. 

In conclusion, the analysis of particular realizations of the active
and passive layers, carried out in the present paper, illuminates the
physics underlying the strong drag in the localized regime.  The {\em
  discreteness} of the hopping electrons gives rise to Coulomb
blockade, which, in turn, opens the possibility of classical pumping.

\end{multicols}
\end{document}